\documentclass[letter]{aa}
\usepackage{graphicx}
\usepackage{afterpage}
\usepackage{here}
\usepackage[varg]{txfonts}
\usepackage{natbib}

\begin{document}

\title{Molecular absorption in transition region spectral lines}
\author{D.J. Schmit \inst{1}
\and D. Innes \inst{1}
\and T. Ayres \inst{2}
\and H. Peter \inst{1}
\and W. Curdt \inst{1}
\and S. Jaeggli\inst{3}}

\institute{ Max Planck Institute for Solar System Resarch
\and Center for Astrophysics and Space Astronomy, University of Colorado
\and Department of Physics, Montana State University}

\abstract{}{We present observations from the Interface Region Imaging Spectrograph (IRIS) of absorption features from a multitude of cool atomic and molecular lines within the profiles of Si IV transition region lines. Many of these spectral lines have not previously been detected in solar spectra.}{We examined spectra taken from deep exposures of plage on 12 October 2013. We observed unique absorption spectra over a magnetic element which is bright in transition region line emission and the ultraviolet continuum. We compared the absorption spectra with emission spectra that is likely related to fluorescence.}{The absorption features require a population of sub-5000 K plasma to exist above the transition region. This peculiar stratification is an extreme deviation from the canonical structure of the chromosphere-corona boundary . The cool material is not associated with a filament or discernible coronal rain. This suggests that molecules may form in the upper solar atmosphere on small spatial scales and introduces a new complexity into our understanding of solar thermal structure. It lends credence to previous numerical studies that found evidence for elevated pockets of cool gas in the chromosphere.}{}
\maketitle
\section{Introduction}
Multiwavelength studies of the solar atmosphere have revealed that the temperature structure is highly non-uniform.
The balance of advection, conduction, radiation, and mechanical and electromagnetic heating produce the transition from photospheric to chromospheric to coronal thermodynamic conditions.
Additionally, the thermal state of the solar atmosphere varies strongly based on the strength and topology of the magnetic field, which produces a highly anisotropic atmosphere in all three spatial dimensions.
While previous authors have constrained the canonical model for the time-independent stratified atmosphere \citep{val_2,val,fontenla_93}, subsequent studies have revealed that in reality the solar atmosphere is likely in a constant state of flux with rapid and localized departures from the unstable VAL atmosphere \citep{{ayres_81},{mcclymont_83}, {gudiksen_11},{carlsson_97}}.
The IRIS mission \citep{depontieu_14} has been designed to study these dynamics with unprecedented resolution and cadence.

Because of the gradients in temperature and pressure, molecular populations are believed to be localized in a narrow region surrounding the temperature minimum.
Carbon monoxide and molecular hydrogen are believed to be the most populous species in the chromosphere, based on abundances and formation channels.
In general, molecular spectral lines are present as absorption features in the optical and infrared \citep{grevesse_73} because of their cool formation region, but instances of emission spectra through fluorescence processes have also been observed \citep{jordan_78,bartoe_78}.
In this paper, we present observations of previously undocumented molecular lines in the Sun within the spectral range from 1390\AA~to 1405\AA.
These lines are seen in absorption within the emission profile of Si IV 1394\AA~and 1403\AA~which form at temperatures an order of magnitude higher \citep{avrett_13}.
These peculiar features are seen near a patch of weak field within the plage and are not associated with a sunspot or pore.
These observations add new constraints to the increasingly complex picture of the temperature and morphological structure of the magnetic chromosphere.
A subsequent paper will focus on diagnostics, whereas this work serves to introduce these important phenomena and their context.
\begin{figure*}
\centering
\includegraphics[width=.9\textwidth,trim=0mm 10mm 0mm 10mm]{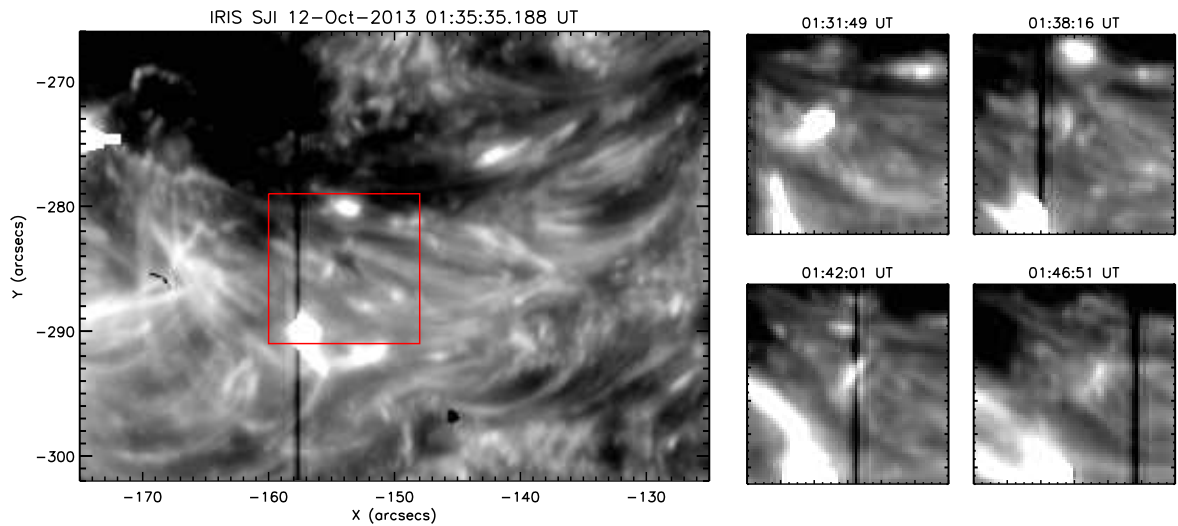}
\caption{IRIS slit jaw images in the 1400\AA~bandpass of the plage region on 12 October 2013. The red box surrounds the region of molecular absorption. The slit position is given by the vertical black line at X=162".  The four small frames show the boxed region at alternate times. The color table is logarithmic. An animated version of this figure is presented in the online journal.}
\end{figure*}
\section{Method}
The IRIS spectrograph observes the solar atmosphere in three spectral windows in the ultraviolet, focussing on Mg II (2800\AA), C II (1330\AA), and Si IV (1400\AA) lines.
A complete description of the instrument can be found in \cite{depontieu_14}.
Our dataset was obtained using the OBS ID 3880013646.
This observing mode outputs full spectral resolution of the full detector (on the three spectrograph CCDs) using 0.352" raster steps and 30 second exposures.
The observation includes a single raster which began at 11 Oct 2013 23:54UT and ended at 12 Oct 2013 03:29UT.
The raster region spans the plage surrounding sunspot group 11861.
This was an active sunspot group with many mid-C class flares spanning the 24-hour period before and after our observations.
Our analysis has been conducted using Level 2 data (served at iris.lmsal.com) which has been flat fielded and dark current subtracted by the methods described in \cite{depontieu_14}.

\section{Results}
We studied spectra from a magnetic element located at Solar-X=-153", Solar-Y=-284".
This magnetic feature is of the same polarity (negative) as the nearest sunspot.
It is bright in Mg II, 1390\AA~continuum, and the bandpass images at 1600\AA~and 1700\AA~of SDO/AIA \citep{lemen_12}.
Figure 1 shows this region as seen in the IRIS slit jaw camera using a bandpass filter to image the spectrum around 1400\AA~(FWHM=50\AA).
An animated version of this feature appears in the online version of the journal.
At 01:35 UT, a dark flame-shaped protrusion is visible extending from the magnetic element.
In the movie, this dark material is visible in a few frames before the spectrograph slit overlays the site at 01:42UT.
The material is most likely dark because of absorption, and it appears to disperse between 01:36 UT and 01:43 UT.
Neighboring the dark material to the east is a source of variably strong Si IV emission.
While large coherent loops are visible in the slit jaw movie to the far east of the magnetic element, we do not see any clear loop structure emanating from the magnetic element.
This suggests that the Si IV emission is either located at the footpoint of a loop that extends into the upper atmosphere and reaches coronal temperatures or composed of many small loops (radius $<200$ km) that are spatially unresolved.

The upper panels of Figs. 2 and 3 show the spectra taken at 01:42UT when the slit is over the magnetic element.
An animated version of the spectra is available in the online version of the journal.
The strongest emission lines visible along the slit are the 1393.8\AA~and 1402.8\AA~Si IV lines, while the O IV lines at 1399.8\AA~and 1401.2\AA~and the Mg II line at 1398.8\AA~are significantly weaker.
The unique absorption pattern is most clearly visible between Solar-Y of -283.5" and -284.5".
The Si IV profiles are broadened by a factor of 3 beyond their nominal width and the continuum is enhanced by 20\% relative to the surrounding plage.
These characteristics are normal for small magnetic concentrations.
However, within the core of the Si IV lines, rather than a smooth Gaussian, the line profile is serrated with many local minima and maxima.
While these profiles could be related to Doppler shifts and unresolved flows, we find substantive evidence that much cooler plasma is producing a complex absorption pattern throughout the spectra between 1330\AA~and 1405\AA.
Under close examination, we find that the continuum surrounding Y=-284" contains many dark lines (see Figure 3).
Based on the values cited in \cite{depontieu_14}, we estimate a signal-to-noise ratio of five in the continuum at 1400\AA.
Ascertaining the significance of the profile structures within the Si IV lines is a difficult task given the complex profiles that are regularly observed in plage with IRIS.

We cross-check our potential absorption features against emission spectra observed several raster steps later at 02:41 UT, which is shown in the lower panels of Figs. 2 and 3.
A slit jaw image that shows the region surrounding the molecular emission spectra is presented in Figure 4.
The emission lines are observed in a weak-continuum, weakly emitting Si IV area but are near a flare.
We believe that these lines are related to a fluorescences process.
Lyman-$\alpha$, O IV, and Si IV resonance lines have all been tied to molecular fluorescence \citep{jordan_78}.
Based on the emission spectra, we identify 13 potential lines in and surrounding Si IV 1403\AA, 12 lines in Si IV 1394\AA, and 11 lines in the continuum.

We have several clues that have allowed us to partially identify these lines.
The close spacing of the lines suggest a molecular origin as electronic-vibrational transitions often occur in clustered sets.
Many of the lines are narrow.
At cool temperatures, at or below the quiet sun temperature minimum suggested by \cite{val_2}, molecules can readily form and thus we expect a sudden increase in line formation.
These cool temperatures are confirmed by the positive matches for atomic lines in the absorption spectra: Ni II 1393.3\AA, S I 1401.5\AA, and Fe II 1403.1\AA.

Molecular lines have been observed in the solar spectrum as absorbing features in the optical and infrared continuum \citep{grevesse_73}.
These lines aid in diagnosing the structure of the photosphere and temperature minimum. 
Sunspot spectra have shown some molecular lines in emission, primarily identified as CO and H$_2$.
It is believed that fluorescent pumping (from above) in transition region lines excites these lines \citep{jordan_78,bartoe_78}.
Neither of these scenarios fits our observation: absorption within transition region lines.

\afterpage{
\begin{figure*}
\centering
\includegraphics[width=.9\textwidth]{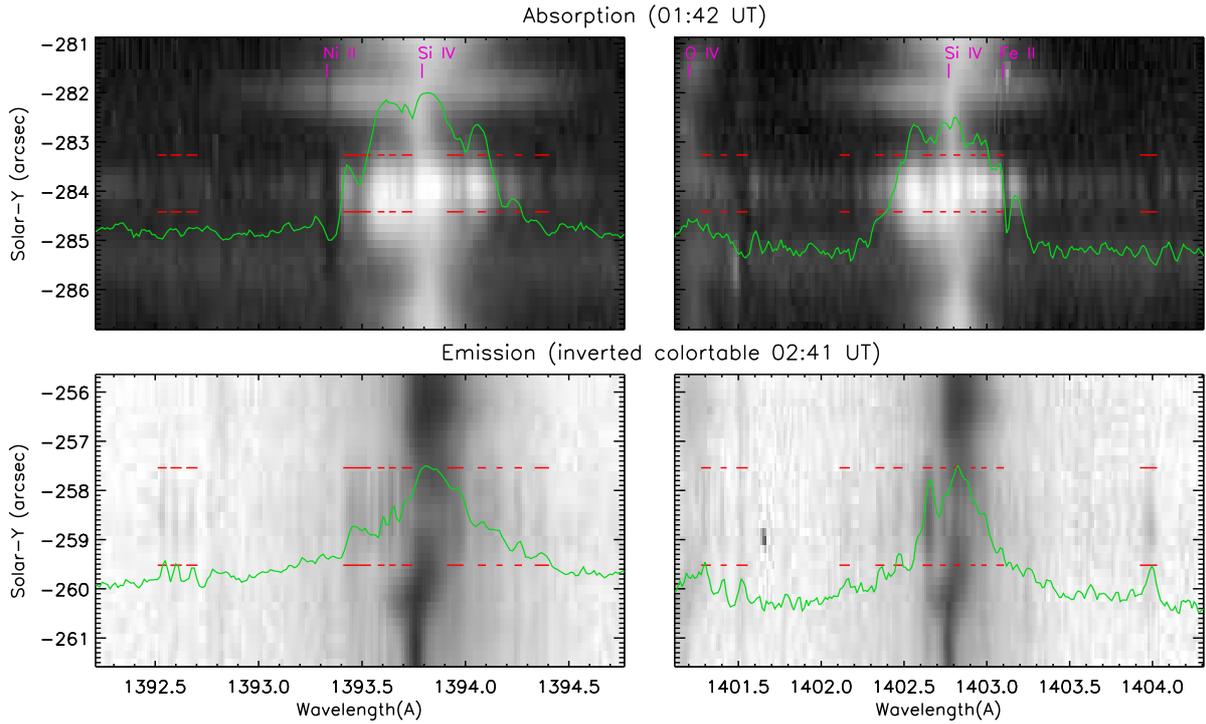}
\caption{IRIS spectra surrounding the Si IV lines at 1393\AA~and 1402\AA~using a 30s exposure. Upper panels show the absorption spectra taken at 12 October 2013 01:42 UT. The lower panel uses the fluorescent emission spectra with an inverted colortable (bright lines are black). The horizontal red lines show the location and width of molecular emission lines and are overlaid on the absorption spectra. The green profiles show the spectra (median filtered along the slit) of the region of the interest. The color table is logarithmic. An animated version of the absorption spectra is presented in the online journal. Each raster step moves 0.35" westward.}
\end{figure*}
\begin{figure*}
\centering
\includegraphics[width=.9\textwidth]{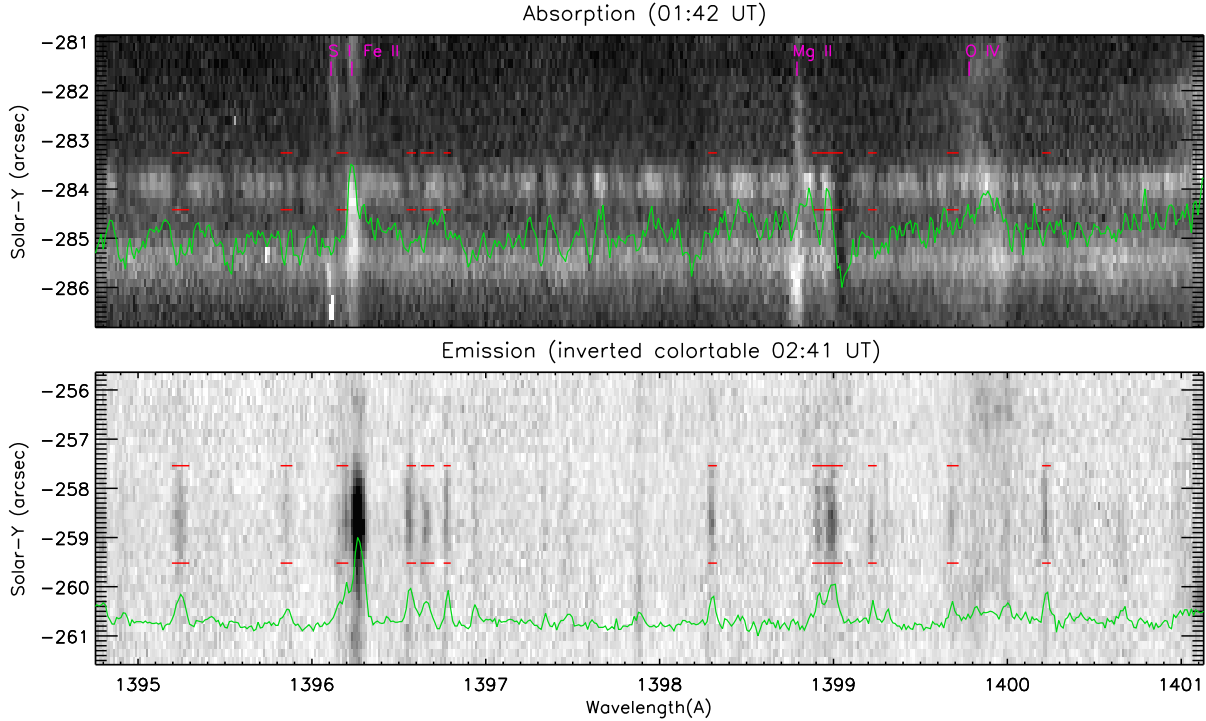}
\caption{Same as Figure 2 but showing the continuum between the Si IV lines. The color table is linear.}
\end{figure*}
\clearpage
}

We have compared our spectra with molecular transition tables compiled by Robert Kurucz (served at http://kurucz.harvard.edu/linelists/linesmol/; see \citealt{kurucz}) for H$_2$ and CO.
There are 360 CO A-X band (including 27 A5-X0) electronic transitions with rotational-vibrational splitting that sit within 700 m\AA~ of the Si IV 1402.8\AA~line and 358 CO A-X band (including 22 A5-X0) transitions within the profile of Si IV 1393.8\AA.
There are 24 H$_2$ B-X band transitions in Si IV 1402.8\AA~and 35 in Si IV 1393.8\AA, but these lines have excited lower levels ($>2\times10^4$ cm$^{-1}$)  which makes them less likely absorbers at 4000K.
Given the number of transitions and the broad range of excitation energies, we would not expect the fluorescence emission and absorption spectra to have identical signatures, thus we do not find a perfect correlation between the datasets.
The unresolved blending of the these transitions can explain the variety of widths in the line profiles.
Although these lines have not been observed in absorption in the Sun, they have been documented in other stars \citep{ayres_03}.

In order to properly identify the contributing transitions, a complete radiative transfer model must be used to incorporate the effects of both atoms and molecules.
This modeling effort is beyond the scope of this letter, but the authors are now at work on this study.
Our subsequent paper will use this radiative transfer model to diagnose the physical conditions of the multithermal plasma along the line of sight.

\afterpage{
\begin{figure}
\includegraphics[width=.5\textwidth,trim=0mm 10mm 0mm 10mm]{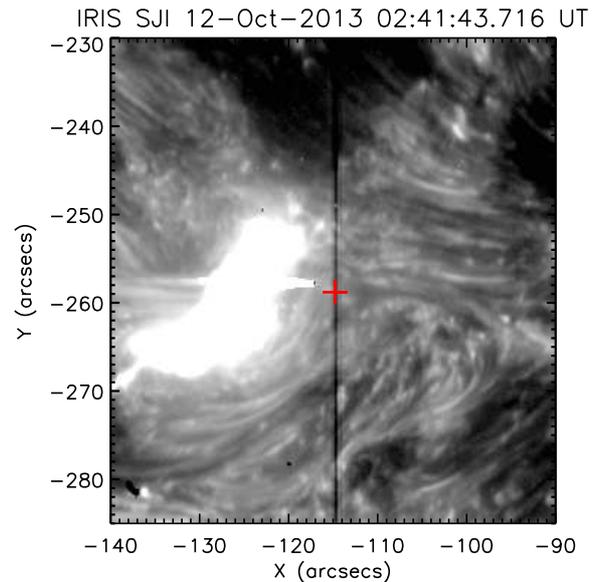}
\caption{IRIS 1400 \AA~slitjaw image where the red cross indicates the location of the molecular emission. The color table is logarithmic.}
\end{figure}
}

\section{Discussion}
An unusual physical model is required to explain these observations.
We observe cool (photospheric temperature) plasma above a source of Si IV emission.
Sadly, the limits of the dataset make a complete model difficult to constrain.
We only have a single raster of this region so we are not able to disambiguate the spatial effects from the temporal.
The 1400 \AA~slit jaw data show that the source region of the absorption spectra is strongly variable in emission over the three-hour observing period.
The spectra are observed during a time of strong brightening with a duration of 25 minutes.
Given the enhanced Si IV widths, it is possible that the spectra are related to an explosive event, but the brightening duration is significantly longer than those observed by \cite{innes_97}.
The proximity of the dark structure shown in Figure 1 (prior to the slit overlapping the region) offers a possible candidate for the source of the cool material.
However, it is difficult to extract an accurate structural morphology from the slit jaw images.
Is the dark plasma a cloud (of ejected cool plasma or cooling coronal plasma) that sits along canopy magnetic field lines well above the photosphere?
Or are we observing low lying cool material near a loop footpoint which has a deeply embedded transition region (the buried corona of \citealt*{ayres_03})?
The simulations reported by \cite{leenaarts_11} suggested that the propagation of shocks through the non-magnetic chromosphere could produce pockets of high-altitude molecule-rich plasma.

The ratio of O IV 1401.4 \AA~and Si IV 1402.8 \AA~can be used as a density diagnostic to probe the source region of the transition region plasma.
Collisional de-excitation becomes important at high densities for the O IV 2p$^2$ $^4$P state, thus the intensity relative to the Si IV 1402.8\AA~resonance line scales inversely with density (H. Peter private communication).
Based on CHIANTI data \citep{dere_97}, we estimate the electron density of the Si IV forming region between $10^{12}-10^{13}$ cm $^{-3}$ which is chromospheric, not photospheric, in magnitude.
The Mg II transitions may aid in our diagnosis of the physical state of the plasma.
The subordinate Mg II 2790\AA~and 2797\AA~lines exhibit similar profiles to the 2795\AA~and 2802\AA~(k and h) lines: broad wings, reversed core, approximately equivalent integrated intensities.
The Mg II 3d $^2$D state is not normally populated because of its high excitation energy and magnetic dipole transition to the ground state.
The strong subordinate line emission may be an indicator of recombination, which would imply that our molecular absorbers are related to a dynamic cooling process.
A complete diagnosis requires a holistic approach of considering the density and temperature along the line of sight and solving for the radiative transfer of the Mg II, Si IV, H$_2$, and CO lines and continuum at 1400\AA~and 2800\AA.
Our upcoming study will take this approach, and further elucidate the details of the small-scale structure within the chromosphere.

\bibliographystyle{aa}


\end{document}